\documentclass[fullpage,epic,eepic,epsfig,psfig,12pt]{article}
\usepackage{graphicx,epsfig,psfig,fullpage}
\usepackage[numbers,round,sort]{natbib}
\usepackage{fullpage}
\makeatletter \renewcommand\@biblabel[1]{${#1}.$}
\makeatother

\begin{document}

\Large
\Large
\centerline{\bf TransPath: A Computational Method to Study the}
\centerline{\bf Ion Transit Pathways in Membrane Channels}

\normalsize
\vspace{1. cm}
\centerline{Zhifeng Kuang,$^{a}$ Anping Liu,$^{a}$ and Thomas L. Beck$^{a,b,}$
\footnote{To whom correspondence should be addressed: becktl@email.uc.edu}}
\centerline{\it Department of Chemistry$^{a}$}
\centerline{\it Department of Physics$^{b}$}
\centerline{\it University of Cincinnati}
\centerline{\it Cincinnati, OH 45221-0172}
\centerline{\today}

\vspace{1. cm}
\large
\centerline{\bf Abstract}
\vspace{.3 cm}
\normalsize

The finely tuned 
structures of membrane channels allow 
selective passage of ions through the available 
aqueous pores. In order 
to understand channel function, it is 
crucial to locate the pore and study its 
physical and chemical properties. 
Recently obtained X-ray crystal structures of 
bacterial chloride channel homologues
reveal a complicated topology with curvilinear pores. The commonly 
used HOLE program encounters difficulties in studying such pores.
 Here we propose a new pore-searching 
algorithm (TransPath) which uses the Configurational 
 Bias Monte Carlo (CBMC) method to 
generate transmembrane trajectories driven by both
geometric and electrostatic features. The trajectories are binned 
into groups determined by a vector distance criterion. 
From each group, a representative trajectory is 
selected based on the Rosenbluth weight, 
and the geometrically optimal path is obtained by simulated annealing. 
Candidate ion pathways can then be  
 determined by  
 analysis of the radius and potential profiles.
 The proposed method and its implementation are illustrated using the 
 bacterial KcsA potassium channel as an example. The procedure is 
then applied to the more complex structures of the bacterial {\it E.~coli} 
 ClC channel homologues.

\newpage
\section{Introduction}

Ion channels comprise an important class of membrane proteins which control the selective
passage of ions into and out of the cell \citep{Hille01}. They possess aqueous 
pores through which the ions pass when the channel is in the open state. 
Many physiological functions in living cells are regulated  
 by selectively translocating ions through a channel pore.   
The study of the physical and 
 chemical properties of the pore provides clues concerning the function of 
an ion channel.  These insights can aid in motivating site-directed mutagenesis
studies and new drug design. 
With the advent of complex channel structures from X-ray crystallography, 
it is desirable to develop an effective computational tool to 
locate and display the pore, analyze the 
dimensions of the pore, identify the pore-lining residues, 
and produce the potential energy profile along the pore.

Several methods have been developed to display and analyze the surfaces of
and cavities in proteins.  These methods include the Connolly surface 
method \citep{Connolly83}, 
VOIDOO \citep{Kleywegt94},
PROACT \citep{Williams94}, and SURFNET \citep{Laskowski95}. A specially 
 designed program, HOLE \citep{Smart93,Smart96}, has been used for 
 more than a decade to provide information
 about the variation of radius along a channel pore. This method is 
useful for relatively straight pores such as the gramicidin A and potassium 
channels. For complex structures 
like the ClC Cl$^-$ channel homologues, however,
 it is difficult to use this  
  method to locate and 
display the ion conduction pore due to the curvature feature, multiple pores,
and possible blockage of the pore by residues 
in the protein.  

 The basic idea of the HOLE algorithm is to use a Monte 
Carlo simulated annealing procedure to find the largest radius of a probe 
sphere which can be accommodated along the length of the pore.  HOLE moves the probe by 
taking only geometry into consideration. When there are multiple geometric pores, it 
is difficult for HOLE to distinguish an ion permeation pore from other energetically
inaccessible pores during the hole searching process.
 In addition, HOLE requires 
a user to specify an initial point within the pore center and a direction 
of the pore alignment.  
 That means that HOLE is 
 specially designed 
 for analyzing the dimensions of a known pore but not for searching for unknown pores.
 Finally, HOLE uses $\exp(R/K)$ as its importance sampling function, where $R$ is 
the distance from the center of the probe to the edge of the nearest protein 
atom and K is a control constant analogous to the factor $kT$ in 
equilibrium statistical mechanics. 
Due to the soft nature of the `energy' $-R$, the procedure allows the probe 
to move through highly confined or even occupied regions of space. 
While the HOLE algorithm has limitations for complex topologies like the ClC 
Cl$^-$ channel homologues, Monte Carlo (MC), molecular dynamics (MD), and Brownian dynamics (BD) 
simulation methods have been employed to study the ion permeation 
process \citep{Miloshevsky04,Cohen04,Corry04,Berkowitz04,Chung05}.
 It is very time consuming, however, to calculate the full interaction energy for each 
step of ion motion.\footnote{Other coarse-graining approaches have been developed
to significantly extend the accessible timescales in studies of ion permeation
\citep{Elber03}.} 
 
In order to address these issues, we propose a new method to 
  generate transmembrane trajectories by taking advantage of the 
Rosenbluth self-avoiding random 
walk scheme 
\citep{Rosenbluth55,Siepmann92}.
The biased random walk approach 
allows for
moves in the configuration space which may take place only over very long times 
in MD simulations.  
With the incorporation of continuous
potential energy terms in the sampling function, the approach is termed the
Configurational Bias Monte Carlo (CBMC) method, which has found wide application
in polymer simulations \citep{Siepmann99,Frenkel02,Beck05}.  
In our work, the sampling function
includes contributions from both geometric and electrostatic features in the channels.
These terms direct the biased random walk through open regions of space that are
energetically favorable for ions with a specified charge.  

Following the generation of an ensemble of membrane-traversing trajectories,
we then utilize a simulated 
annealing technique similar to the HOLE method to find a unique and geometrically optimal
path for analysis. 
The simulated annealing 
procedure is used to adjust each state to the center at which  the largest 
sphere can be accommodated without overlapping any protein atoms.
We utilize a harsher sampling function than is used in the HOLE method in order
to prevent movement into highly restricted domains.   
Upon annealing, the trajectories typically collapse into one
or a few protopaths. 
 Finally we calculate the radius and potential profiles for each of these 
protopaths.  
By examination of both the radius and potential profiles, 
candidate ion conduction paths can be identified.  This method has previously been
employed in a detailed study of ion transit paths and gating in the 
bacterial ClC channel homologues \citep{Yin04,Yin05}; here we present details of the algorithm.
In our previous work, we focused on the extracellular side of the structures to explore
chloride ion pathways, possible binding sites for chloride ions awaiting entry to 
the filter, multi-ion occupancy of the filter, proton access 
pathways approaching the fast gate, and the dependence of glutamate gate protonation
on occupancy of the central region of the filter by chloride ions.  In the present study
we examine electrostatic effects on the intracellular side of the 
chloride channel homologues and
relate our results to recent, related physiological experiments \citep{Chen03,Chen03a}. 

In the Methods section, 
 we describe the algorithm by taking the KcsA bacterial 
 potassium channel as an example. The K$^+$ channel is a 
 tetramer with four-fold symmetry about a central pore \citep{Doyle98}. 
From the intracellular side, the pore begins as a tunnel and then opens into a wide cavity. The other end 
of the cavity is connected to the narrow filter which separates the cavity 
 from the extracellular 
solution. This well-defined pore has previously been analyzed with the HOLE method 
\citep{Doyle98,Biggin02,Ranatunga01}, so it 
 serves as a benchmark for our algorithm.  
 In the Results Section, we apply 
the method to the more complex {\it E.~coli} ClC Cl$^-$ channel homologues.
 The computational results are compared with experimental 
 observations in the Discussion section.  
 
\section{Methods and Implementation}

We numerically solve the variable-dielectric Poisson equation and thus
generate the electrostatic potential due to the protein in the 
inhomogeneous dielectric environment.  
The problem is discretized on a cubic lattice.  
 The lattice size is 0.5~{\AA}. There are 257 mesh points in one 
 direction, so the length of each side of the cube is 
128~{\AA}. The center of the cube is at the origin of  
 a Cartesian coordinate system $(x, y, z)$.  
 The 3D structure of the protein is discretized with the CHARMM PBEQ   
subroutine \citep{Brooks83} and mapped  
 onto the center of the lattice.  The $z$ direction is perpendicular to the membrane,
passing from the intracellular side (negative $z$) to the extracullular side (positive $z$)
of the protein. 
 The mass center of the protein coincides with  the center of the 
cube.  
 The Cardinal B-spline method is used for charge partitioning. 
 The protein surface is created with a probe radius of 1.0~\AA~for 
the dielectric boundary.
The protein is embedded in a dummy membrane modeled as a uniform dielectric 
medium without any charges or dipoles. 
 The 
lattice is partitioned into three regions. The first 
region  
represents the protein and  
 is characterized by a low dielectric constant 
$\epsilon_p=4.0$.
 The second region includes the solvent 
 reservoirs 
and all the  
cavities of the protein and is characterized by a high 
dielectric constant 
$\epsilon_w=80.0$.
 The third region represents the membrane and is 
 characterized by a low dielectric constant $\epsilon_m=2.0$.  
We set the electrostatic potential to zero on the distant boundaries,
and the potential is calculated with our fast multigrid Poisson solver \citep{Beck00}.
The salt concentration is set to zero for the work presented here.

 As we have already mentioned, the first step of our method is to 
 generate transmembrane trajectories 
 of a probe ion through the channel by use of the CBMC scheme. 
 For the simple case of 
hard-sphere lattice polymer simulations, 
a chain configuration is generated by first inserting a segment at a randomly 
selected site, and then inserting subsequent segments of the chain randomly 
at other available nearest-neighbour sites which are not occupied either by 
other molecules in the system or by any previously inserted segments. In the 
case that no free neighbour sites are available, the attempt to grow a trial 
chain configuration is abandoned. 
 In this way, a continuous hole in a protein can be probed, so we suggest that 
 the self-avoiding random walk (SAW) scheme of Rosenbluth and Rosenbluth \citep{Rosenbluth55} can 
be modified to search for the ion conduction pathway of a channel based on its 
X-ray crystal structure. The additional feature in the CBMC method is to employ
a biased random walk where the Monte Carlo choice is based on the relative
energetics for possible atom placements (in addition to excluded volume effects).
Below we discuss the algorithm in 
detail. 

\vspace{.5cm}
\noindent
{\bf Step 1. Find the first segment}

We do not insert the first segment at a randomly selected site. Instead, 
we take advantage of the different dielectric constant assignments for 
 the protein ($\epsilon_p=4.0$) and cavity ($\epsilon_w=80.0$) regions to locate all the 
high dielectric spots in a starting window bounded by $[xl,xr]\times[yl,yr]
\times[zl,zr]$. The $z$ window is typically narrow and is chosen near the center of the protein, 
and the $x, y$ windows cover most of the protein in the plane parallel to the membrane.
Since any ion pathway must pass through aqueous zones, an exhaustive search of all
of the high-dielectric patches must contain the relevant pores.
For each located high-dielectric spot, a 
simulated annealing process described below is performed to find the center at which the biggest 
sphere can be accommodated without 
overlapping with any protein atoms, and that high-dielectric spot is not revisited in searching 
for initial locations for trajectories. 
For the KcsA potassium channel, we chose $[-5, 5]\times[-5, 5]\times[-0.5, -0.5]$
 as the starting window, a simple square near the center of the membrane. The first segment  
found is at $(0., 0., -0.5)$. 

\newpage
\noindent
{\bf Step 2. Grow the chain}

 We generate a chain by use of a self-avoiding 
and biased random walk of a probe 
ion of charge $q$ starting from one of the 
 recorded first segments.  The possible sites 
 of a trial step  
 are not 
 neighbour lattice sites but as many randomly chosen sites 
on the surface of a sphere as desired. Hence, the motion is in continuous space.  
At the current step of the random 
walk, we bias the choice of 
the next step by comparing 
 the Boltzmann weights (Eq.~\ref{BIAS} below) of 20 random positions uniformly 
 distributed on a $5/8$ sphere surface centered at the current position and 
pointed in the $z$ direction. The sphere radius is chosen as 
0.25~{\AA}. 
We select the next position $j$ with a 
probability 
\begin{equation}
P_j = \frac{e^{-{E_j}/K}}{\sum\limits_{i=1}^{n}e^{{-E_i}/K}} ,
\label{BIAS}
\end{equation}
where $n\leq 20$ is the total number of 
available sites excluding those positions which are 
 occupied by  protein atoms or previously generated segments. The
energetic factor $E_j$ is specified below, and $K$ is a control constant
analogous to a temperature. 
The product of the denominators in Eq.~\ref{BIAS} for each added segment 
is termed the Rosenbluth weight for the fully grown chain.

Since ion motion across cell membranes has important contributions from the 
physical dimensions of the pore,  the electrostatic potential due 
to the protein, and the membrane potential, 
we propose the following energetic cost term for the chain growth:
\begin{equation}
E_{j} = 
 \omega_r /R + \omega_{\phi} q\phi + 
\omega_{bias} q\phi_{bias} ,
\label{BLTZ}
\end{equation}
where 
$\phi$ is the electrostatic potential due to the fixed protein charge distribution  
and $\phi_{bias}$ is    
 an imposed extra potential which accounts for the membrane potential and 
nudges the biased random walk either up or down. 
 $\omega_r$, $\omega_{\phi}$, and 
$\omega_{bias}$ are scaling factors 
  which can be adjusted.
We have taken $\omega_r = 0.1$, $\omega_{\phi}=1.0$, and
$\omega_{bias}=1.0$ to yield comparable magnitudes for all
three terms. We have
found that, in the generation of the transmembrane trajectories, it is better to
maintain a relatively small value for $\omega_r$; the test ion motions are excluded from
occupied regions but are otherwise largely driven along the biased random walk by
electrostatic features.  
 $R$ is the distance from a trial position ${\bf p}$ to the nearest van der 
Waals protein contact, which 
 accounts for the physical dimension 
of the hole:
\begin{equation}
R=\min\limits_{i=1}^{N}[|{\bf x}_i-{\bf p}| - r_i] ,
\end{equation}
where ${\bf x}_i$ is the position of a protein atom $i$ or the 
 previously occupied position of the probe ion, and $r_i$ is the van der 
Waals radius of the atom or the probe ion.  Our geometric contribution, 
$\omega_r/R$, is proportional to the Born desolvation cost of moving
an ion from an aqueous region into a spherical aqueous pore of radius
$R$ embedded in a large low-dielectric environment.

We note that, in a continuum dielectric model, the ion motion should include
a self-energy term in addition to 
 the electrostatic potential due to fixed protein charges. 
 Thus, the self-energy 
barrier due to the penalty of dehydration is missing in the energy  
function Eq.~\ref{BLTZ}. If this term is included, the self energy must be 
calculated for each trial position, which would be computationally expensive.
The self energy can be quite large in the narrow selectivity 
filter regions \citep{Yin04}.
One justification for neglecting the self-energy contribution is that, during 
the CBMC trajectories, we do already include 
a (relatively small) geometric term 
which directs the 
walk into open spaces, and this is related to a simple Born expression for desolvation 
in a spherical cavity (above).  
A second justification for the neglect of the self-energy term relates to 
small protein fluctuations about the ions in narrow regions of the pore.  
It has been shown that these fluctuations, while not appreciably altering the 
self-energy contribution, can lead to a large electrostatic stabilization 
energy which can nearly offset the self-energy penalty \citep{Mamonov03}.  All the calculations
presented here are for rigid protein structures which don't include this stabilization
effect countering the self energy.  Thus, neglecting the self energy approximately 
accounts for this partial cancellation. 

The above 
procedure is repeated step-by-step until either    
(a) the trajectory has walked out of the protein, 
or (b) it is blocked somewhere inside the protein. 
Those trajectories which 
do not make their way out of the protein are discarded.   
In turn, the whole process is restarted from the initially specified first 
segment  
 by directing the $5/8$ sphere surface in the opposite direction and 
reversing the external potential.   
The process of the CBMC random walk from the same starting 
point is repeated many times. Swarms of transmembrane trajectories are 
 generated and recorded. 

At this stage, ideally we should perform a simulated
annealing process on all transmembrane 
trajectories to see if those starting
from the same first segment will collapse into one or a few prototype paths.
But we find this is expensive and not necessary. A more efficient way is
to group the trajectories by comparing the vector distances between them as follows.

We choose a number of lattice layers
perpendicular to {\it z}
with equal distances of separation of 0.5~{\AA}. The exact number and positions of the layers
depend on the size of the
protein and are automatically adjusted by the code.
 Each trajectory is represented by a vector where the elements 
are the intersection points between the trajectory and the layers.
 
We then take the vector of the first trajectory as the reference vector of
the first group. Subsequently, we compute the distance between the reference vector
and the second vector. If the distance is less than a given cutoff
 distance,  we place it
  in the same group and take the average as a new reference vector;
otherwise, a second group with its own reference vector is created.
Screening all the vectors,
we bin all trajectories into no more than 10 groups. 
From each group,  we choose the
trajectory with maximum Rosenbluth weight as the
 optimal trajectory. 
The averaged Rosenbluth weight is related to the excess
chemical potential, and thus provides a sensible measure of favorable energetics for 
a given trajectory. 

\vspace{.5cm}
\noindent
{\bf Step 3. Simulated annealing}

 Finally, 
a Metropolis Monte Carlo simulated annealing  
procedure is used to adjust each point {\bf p} along the selected trajectory 
 to find 
the largest sphere whose center lies on the plane through  {\bf p} and 
 orthogonal 
 to the {\it z} direction \citep{Smart93}. Since the radius of a pore should be larger 
than 0.5~{\AA}, but the distance between two adjacent states of a trajectory is 
set to be 0.25~{\AA} (the sphere radius for generating trial points), 
it is not necessary to implement the 
annealing procedure for each point on the trajectory.   Rather, we perform the  
 annealing on the corresponding vector obtained in the binning process
described above. 

Although we use a procedure similar to Smart {\it et al.} \citep{Smart93}
for the annealing step,
 there are three features which differ. First, a 
 potential profile along the center is also generated during
 the annealing process.  This potential profile is the basis for
identifying energetically favorable ion conduction pathways.
  Second, we use $1/R$ rather than $-R$ as the geometric `energy' term; if $R$
is negative, the energetic cost is taken as infinity which prevents the paths
from `tunneling' through occupied regions. Given a 
transembrane trajectory generated by the CBMC algorithm, this allows the annealing
process to relax locally to the geometric center without moving large distances
into neighboring geometrically accessible cavities. Third,  
the annealing procedure is performed in planes normal to {\bf z}, whereas 
in the HOLE algorithm the annealing is done in planes perpendicular to the 
chosen search direction.  Otherwise, our algorithm is identical to that in
Ref.~\citep{Smart93}, and the reader is referred there for details.  
The annealing step 
locates the geometric center for the trajectory produced in the CBMC procedure
discussed above.  

We choose the initial annealing control
constant $K$ to be very small (0.001).  In this case, the random walk 
goes steadily downhill since only moves that increase the distance $R$
are accepted.  With a relatively small maximum step size (0.2~\AA), this ensures
that the annealing finds a minimum in the energy function (that is, a 
maximum empty sphere size)
near the already generated trajectory.
In addition, we choose to use the hard-core radius as opposed
to the van der Waals radius employed during the CBMC process 
\citep{Smart93}.  The hard-core
radii are typically smaller than the van der Waals radii, and use of the 
hard-core values during the CBMC step leads to an 
excessively `open' structure which 
allows for penetration of the chains into the protein.
Smart {\it et al.} \citep{Smart93} argue, however, that the hard-core radii may give a slightly
better estimate of pore radii which includes some degree of pore fluctuations 
in an average sense.  Therefore, once the transmembrane trajectories 
are generated, 
we employ the hard-core values for the 
final relaxation to the geometric center. 

The generated ${\bf p}_{new}$, $R({\bf p}_{new})$, the electrostatic potential 
$\phi({\bf p}_{new})$, and the index of the closest protein atoms 
to ${\bf p}_{new}$ are recorded following the annealing process. 
The potential 
$\phi({\bf p}_{new})$ at ${\bf p}_{new}$ is calculated by a 
weighted-average interpolation scheme. The potential at a location
not on a grid point is taken as the average from the neighboring 
eight grid points weighted by the relative volume of each sector within the cube.
We shall refer to the functions  
(${\bf p}_{new}(z), R({\bf p}_{new}))$ and 
(${\bf p}_{new}(z), \phi({\bf p}_{new}))$ as the radius profile and 
potential profile of the pore. The radius profile provides information 
concerning the geometry of the pore. The potential profile provides clues 
about the function of the pore. The closest protein atom index 
is a representation of the pore-lining residues. Based on both the potential 
profile and the radius profile, possible ion conduction pathways can be sorted. 

Our TransPath searching on the KcsA channel utilized membrane
and protein dielectric constants of 4 and a water value of 78 to compare
with the results in Ref.~\citep{Ranatunga01}. The structure was taken 
from the X-ray crystal determination in \citep{Doyle98} (PDB code 1BL8). Hydrogens
and missing atoms were added with the CHARMM \citep{Brooks83} utilities, and were
not relaxed. The default charge state was assumed in order to compare with the lowest curve
in Fig.~5 of Ref.~\citep{Ranatunga01}.
The resulting path moves
along a straight line down the center of the channel, as observed 
in Ref.~\citep{Doyle98}.
Fig.~1 displays the radius and potential profiles along the KcsA channel.  The radius
profile obtained from TransPath is superimposable on the corresponding profile 
from HOLE for the same structure.  The potential profile is 
very close to that in Ref.~\citep{Ranatunga01}, except the potential well near
the intracellular channel entrance is not as deep for our path.  This is likely due
to slightly different placement (compared with Ref.~\citep{Ranatunga01})
of missing charged R117 atoms near the mouth of the pore. 
We do not repeat variations of the charge states or the interpretations of the 
results since they appear in the cited literature.   
 In the next section, we apply 
the procedure described above to the more complex bacterial 
ClC Cl$^-$ channel homologues.

\section{Application to the Bacterial ClC Channel Homologues}

Eukaryotic ClC chloride channels control the selective flow of small anions across membranes.
They are involved in such physiological functions as the acidification of intracellular vesicles,
excitability of skeletal muscle, and salt and water transport across epithelia \citep{Estevez02a}. 
They also are involved
in several inherited diseases \citep{Ashcroft00}.  
The channel gating is voltage and pH sensitive \citep{Chen01}, and the chloride ion itself participates in the fast 
gating process \citep{Pusch95,Chen96}. Chloride channels are less selective than cation (sodium and 
potassium) channels, allowing a range 
of anions to pass under electrochemical gradients, and they exhibit multi-ion conduction \citep{Pusch95}.
Prokaryotic homologues of the chloride channels
have been shown to function in the acid response of {\it E.~coli} by creating electrical shunts
to balance the charge when protons are pumped out of the organism 
during exposure in the stomach \citep{Iyer02}.

In addition, it has recently been discovered that the bacterial homologues are not channels but
proton/chloride antiporters \citep{Accardi04b}.  In our previous 
work on the bacterial structures \citep{Yin04}, we
investigated proton access pathways to the gate from the extracellular
side. We also observed an apparent electrostatic 
conduit for protons from the gate to
the intracellular side of the channel homologues.  
Three acid residues on the intracellular side of the bacterial
channel homologue are replaced by basic or nonpolar residues in
the ClC-0 channel, and a homology model (unpublished)
displays no negative-potential
conduit for protons to move all the way through the channel.  
Mutations of the acidic groups
along the proposed proton pathway would provide interesting evidence
as to how protons migrate across the bacterial structures. 
Previous mutagenesis studies, however, have indicated a 
close structural similarity between the prokaryotic and eukaryotic structures, so the 
functional difference
may result from quite subtle changes in the proteins through evolution \citep{Estevez02a,Accardi04b}.

Recently, X-ray crystal structures have been determined for the bacterial chloride channel homologue and 
two mutants \citep{Dutzler02,Dutzler03}.  These structures have confirmed the double-barrelled 
nature of the channels previously indicated by physiological experiments \citep{Miller84}.
The dimeric structures exhibit a highly complex topology, each monomer containing 18 $\alpha$-helices
with a range of orientations relative to the membrane plane.  Conserved sequences define the selectivity
filter; they occur at the N-termini of several helices, creating a strong positive potential at
a central anion binding site.  Computational studies have shown that the binding energy is 
largely due to local interactions and not overall helix dipole effects \citep{Faraldo04}.

The pore of the bacterial structures displays a curved hourglass structure, with a narrow selectivity 
filter separating two vestibules. In the wild-type structure, a glutamate residue (E148) occludes the 
pore on the extracellular side of the filter, while in the two mutant structures (E148A and E148Q) 
the side chain has moved out of the pore. This suggests a gating function for this strategically placed
residue.  Site-directed mutatgenesis studies support this picture \citep{Dutzler03}, 
but experiments utilizing intracellular anionic blocking agents have suggested that 
more extensive conformational changes
may be involved in gating \citep{Accardi03}.  The crystal structures display three anion binding sites:
one at the intracellular entrance to the filter (S$_{int}$), one in the middle of the filter (S$_{cen}$), and one 
near the extracellular end of the filter (S$_{ext}$).  The last site is occupied either by 
the glutamate gate or a chloride ion. In addition, an 
extracellular binding site (S$_{bs}$) has been 
proposed for chloride ions prior to 
entering the filter 
\citep{Chen96,Miloshevsky04,Faraldo04,Yin04,Berkowitz04}; this 
site is near the positively charged residues R147 and R340 and is adjacent to the E148 gate. 

Several computational studies have appeared recently concerning the modeling of the chloride
channels and their bacterial homologues.  These efforts have focused on the ion transit 
pathways \citep{Miloshevsky04,Yin04}, multi-ion occupancy of the pore \citep{Corry04,Faraldo04,Cohen04,Yin04},
the conduction mechanism \citep{Cohen04,Corry04}, 
the current-voltage behavior \citep{Corry04}, pH dependent gating 
induced by chloride ion occupancy at the S$_{cen}$ and S$_{bs}$ sites 
\citep{Yin04,Berkowitz04,Chung05,Yin05}, and interactions of 
blocking agents with the intracellular side of the pore \citep{Moran03}. A wide range of computational
techniques has been employed, including electrostatics calculations and molecular dynamics, Monte Carlo, 
and Brownian dynamics simulations.  Here we apply the TransPath technique described above to examine 
ion transit pathways on the intracellular side of the bacterial chloride channel homologues. We also examine 
electrostatic changes at the S$_{int}$ site due to mutations designed to mimic recent 
site-directed mutatgenesis studies of the ClC-0 channel \citep{Chen03,Chen03a}. 

We first consider the wild-type {\it E.~coli} structure (PDB code 1OTS) in which the E148 residue
blocks the extracellular end of the selectivity filter. This provides a good test of the 
TransPath algorithm since the HOLE method encounters difficulties near the constricted
gate region.  HOLE searches tend to wander off into the protein near the obstructing E148 side chain.  

For the TransPath pore searching, 
the starting region was confined to the domain 
$[-6, 6]\times [-27,-14]\times [-10,8]$ (subunit A). The $z$
domain was chosen over a larger range since
 two chloride ions and E148 were resolved in this region in the X-ray 
structure in subunit A. In addition, there is a binding site S$_{bs}$
adjacent to R147 (and R340) near the extracellular entrance to the filter \citep{Faraldo04,Yin04}.
The parameters used in the searching 
processes were
$\omega_r = 0.1$; $\omega_{\phi}=1.0$; 
$\omega_{bias}=1.0$; $\phi_{bias}= 150mV$; $K=1.0$ for the CBMC searches,
$K=0.1$ for the initial control constant while searching for high
dielectric spots, $K=0.001$ for the inital control constant 
for the geometric annealing step, and $D_{max}=0.2$~{\AA}~(maxiumum 
step size during simulated annealing).
The default charge state was assumed 
(see below).

From the starting domain, sets of successful transmembrane trajectories 
were generated from 12 different high dielectric spots.
In each of the 12 sets, 10 groups were generated, 
and from each group one trajectory was chosen based on the Rosenbluth weight. 
Each of these 120 trajectories was then annealed, and only several 
unique paths were located.  
By examining the radius and potential profiles, two prototypes were found 
favorable for chloride permeation on the intracellular side as shown in Fig.~2a. 
The 
radius and potential
 profiles are shown in Fig.~3.

The two anion transit pathways are identical for $z$ values greater 
than -10~{\AA}.  This region corresponds to the selectivity filter (roughly $-10 < z < 5$~{\AA}) and
the extracellular vestibule ($z > 5$~{\AA}).  The binding site S$_{int}$ is located 
at $z = -8.7$~{\AA}, which is at the origin of the filter on the intracellular side of the protein.
The narrow constriction near the E148 residue of the 
wild-type structure is apparent in the radius profile ($0 < z < 5$~{\AA}; 
$z = 1.0$~{\AA}~for the S$_{ext}$ site) (Fig.~3a). A larger
radius is also observed near $z = -2.7$~{\AA}~where 
an ion is located in the crystal structure at S$_{cen}$. 
It is clear from the radius profile near $z = -6$~{\AA} that protein 
fluctuations are required to allow ion transit from the intracellular 
side towards the S$_{cen}$ site, since the radius is less
than the ionic radius of a chloride ion; the side chain of S107 is located
at $z = -5.5$~{\AA} and contributes to the decreased radius there.  
Both paths pass through the two anion binding sites S$_{int}$ and S$_{cen}$
and near the E148 obstruction at S$_{ext}$.
The potential
profile (Fig.~3b), while positive through the whole filter, is significantly reduced
in magnitude relative to the two mutant (open) structures near the E148 side chain \citep{Yin04}; 
it is large
enough to bind an ion at the S$_{cen}$ site \citep{Faraldo04,Yin04}, 
but the pore is closed to chloride permeation
due to the small radius and less favorable potential near S$_{ext}$.  
Finally, an extracellular vestibular
anion binding site (S$_{bs}$) is apparent as a shoulder in the potential
in the range $7 < z < 11$~{\AA}. 
Two basic residues (R147 and R340) are located in close proximity to this site.  

The two paths differ in the 
intracellular vestibule.  One path (P1) displays a large radius and positive potential
for $z < -10$~{\AA}, while the other (P2) exhibits a reduction of radius and rapid potential
variation.  The path P1 moves through the more open vestibule region, while the path P2
passes near the $\alpha$-helices D and R.  The potential
peak near $z = -19$~{\AA} is predominantly due
to the R120 residue between helices D and E, while the valley near $z = -16$~{\AA} mainly results from
the E111 residue in helix D \citep{Dutzler02}.  These results suggest that the dominant path 
for chloride entrance to the filter is P1 since the P2 path has an unfavorable potential
energy near $z = -16$~{\AA}.  Finally, 
the pore-lining residues of the paths in Fig.~2a include 
 the key residues Glu148, Phe357,
 Tyr445 and Ser107 which define the filter \citep{Dutzler02}. 

For the wild-type (1OTS) structure, the potential at S$_{int}$ is 0.230 V (this values differs
slightly from our earlier work \citep{Yin04} since chain A was used in the present study as 
opposed to chain B previously).  When a chloride ion is added to the S$_{cen}$ binding site, 
the potential drops to 0.114 V .  It has been shown that the residue E113 
is likely protonated due to a potential shift caused by the nearby E203 side 
chain \citep{Faraldo04}.  When we protonate E113, the potential at S$_{int}$ with a chloride 
at S$_{cen}$ becomes 0.178 V (potential shift of roughly 2$kT$ relative to the unprotonated form). 
This shift has been found to be necessary to ensure favorable free energies of binding of a chloride ion 
at the S$_{int}$ site with ions occupying the S$_{cen}$ and S$_{ext}$ sites \citep{Faraldo04}. 
In our previous work, we found a strong electrostatic well for chloride ions at the S$_{ext}$ 
site (mutant structure 1OTU), even when the S$_{cen}$ site was occupied by another anion.  These 
computational results support the picture of multi-ion occupancy of the pore observed 
experimentally \citep{Pusch95,Dutzler02,Dutzler03}.

Recent single-channel measurements have shown a strong effect of the residues E127 and K519
on ClC-0 channel permeation and gating \citep{Chen03,Chen03a}. 
The two residues are believed
to be located in helices corresponding to the helices D and R from the bacterial channel homologues, respectively,
based on sequence alignment \citep{Dutzler02}. The locations are
near the intracellular entrance to the filter.  
The experimental results indicate an 
electrostatic influence, but it was found not to be due to a simple surface charge effect. This 
is likely due to the fact that the S$_{int}$ binding site is at the entrance to the filter
so that charge screening by large non-permeating ions is not easily achieved. 
A two-site kinetic model 
of the data was proposed: equilibrium binding to the S$_{int}$ site followed by passage of the ion
to a central binding site (S$_{int}$) and escape to the extracellular solution.  In this model, 
the rate-limiting step for saturated concentrations 
was proposed to be the final escape to the solution.  Mutations of the E127
and K519 residues in ClC-0 can affect the binding affinity and the rate of passage into the filter
(first two steps of the model).  Here we focus on electrostatic changes that might influence 
the affinity for binding at S$_{int}$ and the rate constant for passage into the filter 
binding site $S_{cen}$. 

E127 from ClC-0 corresponds to E111
in the {\it E.~coli} channel homologue.  K519 in ClC-0 aligns with T452 in the bacterial structure,
however, and I518 aligns with R451.  Therefore, we mutated R451 into I451 and T452 into K452 to 
mimic the expected local structure in ClC-0.  Following the mutation, the side chains 
of the two residues were relaxed. Both the resulting E111 and K452 side chains 
are in close proximity to the anion pathway and the binding site S$_{int}$ 
(Table 1).

For this work we utilized the mutant E148Q structure
(PDB code 1OTU), in which the glutamine side chain is moved away from the S$_{ext}$ 
binding site.  
We assumed the default charge state of E113 for 
these calculations since we are mainly interested in potential shifts due to the mutations 
at the S$_{int}$ site. 
Anion pathways similar to those observed for the two mutants in our previous 
work were found \citep{Yin04}.  A swarm of representative 
trajectories is shown in Fig.~2b, which clearly displays
the two vestibular regions separated by the narrow selectivity filter. 

The mutation of the bacterial structure described above moves a positive charge 
more than 3~{\AA}~closer to the 
S$_{int}$ binding site, and thus causes a positive shift in the potential relative to the original 
bacterial structure (Tables 1, 2).  
The potential shifts from 0.235 V for the 1OTU structure to 0.435 V for
the I451/K452 mutant structure.  This stronger binding energy at S$_{int}$ 
(roughly 7 $kT$ shift) may be one cause for increased conductance in the eukaryotic ClC-0 channel
relative to the bacterial case \citep{Accardi04b}. 

Next, we performed eight mutations on the E111/K452 pair, sampling all possible charge states
at these two sites.  Neutral states were obtained by protonation of the glutamate or deprotonation
of the lysine.  The results for the potential and the shift relative to the E111/K452 pair 
are shown in Table 2.  First, notice that all potentials at S$_{int}$ are
positive, even in the presence of the E-E- mutation.  Experimentally, the mutations 
corresponding to our mutations E-K+ (default ClC-0), E-K0,
E-E-, E0K+, E0K0, E0E-, K+K+, and K+E- were made, and 
single-channel currents were measured \citep{Chen03}.
The wild-type E-K+ case exhibits the largest low concentration conductance, while the E-K0 and E-E- 
mutants show 
decreased conductances.  All three mutants exhibited similar maximal conductances
with increasing intracellular chloride ion concentration.   
This is consistent with an electrostatic effect on 
equilibrium binding at the S$_{int}$ site due
to the ClC-0 residue 519 charge state, where
the decreasing potential (shifts of -0.245 V and -0.350 V) at S$_{int}$ reduces anion occupancy. 

The E0K+ and 
E0K0 ClC-0 mutants display conductance 
{\it vs.}~chloride concentration curves very similar
to the wild-type channel; the E0E- mutant displays slightly reduced conductance at low concentrations,
but the conductance is significantly 
larger than for the E-E- mutant \citep{Chen03}. These experimental findings 
suggest an important electrostatic role for E127 in the ClC-0 channel; the electrostatic effect
of K519 is to some extent mediated by the negative charge on E127. The potential shifts
for the mutations (E0K+, E0K0, and E0E-) are modest in magnitude, the largest value occurring for the
E0E- mutant (-0.224 V or roughly 8 $kT$). The experimental 
conductance {\it vs.}~chloride ion concentration curve for the E0E- mutant
lies between the wild-type curve and the E-K0 mutant; our potential shift calculation does
not explain this difference in the maximal conductance
since the potential shifts for E0E- and E-K0 are quite close.  We note that our calculations
are based on a fixed structure obtained from a simple homology model, and thus do not include
any effect of protein fluctuations or significant relaxation from the bacterial X-ray structure. 

The K+K+ mutant did not conduct \citep{Chen03}; 
the electrostatic calculations yield a potential shift at
S$_{int}$ of 0.629 V (or roughly 23$kT$). This very large positive shift likely binds the anions too strongly, thus
reducing the driving force for ions to enter the filter and subsequently push ions at the S$_{cen}$ site outwards
towards the extracellular vestibule.  The K+E- mutant did conduct, but with a maximal conductance roughly
one half that of the wild-type channel.    This result shows that charges at the E127 and K519 sites
in ClC-0 do not influence conductance symmetrically.  With the side chain placements
in our calculations, we obtained a potential shift for this mutant of -0.053 V, a very small change, 
so our results do not rationalize the reduced maximal conductance observed experimentally.  
The kinetics of ion passage into the filter may be quite complicated and reflect
important contributions from fluctuations; it was earlier proposed that the K519 residue may 
attract chloride ions towards the filter, while the E127 charge provides a push to drive the
ions deep into the filter \citep{Chen03}. All of 
these effects may be mediated by kinetic fluctuations in 
the local structure around the S$_{int}$ binding site.   
 
Therefore, our results shed some light on the electrostatic 
interpretation in Ref.~\citep{Chen03}; 
mutations of the E127 and K519 sites in ClC-0 that induce variations of the 
potential at S$_{int}$ 
with magnitudes less than several $kT$ do not dramatically change the channel conductances.  However,
if the potential change is too large in either the positive or the negative direction, currents
can be significantly reduced or eliminated.  We are not able to rationalize different conductance curves for the 
E-K0 and E0E- mutants which exhibit similar potential shifts at S$_{int}$, nor changed maximal
conductance for the K+E- mutant compared with wild-type ClC-0.  This may be due to the simple homology 
model employed here, to protein fluctuations, or to limitations in a purely electrostatic interpretation.
The results are consistent with the general picture that the residues E127 and K519
can provide electrostatic control of the affinity of chloride ions for the S$_{int}$ 
site and perhaps the kinetics of ion access to the filter.  
Electrostatic control of access to the filter (S$_{cen}$ and S$_{ext}$ sites) via 
a ``foot-in-the-door" mechanism has been implicated in the reduction
of the channel closing rate with increasing intracellular chloride ion concentration \citep{Chen03a} .

\section{Conclusions}

We have presented a new Monte Carlo algorithm for locating ion transit pathways 
through membrane proteins.  The method is robust; it requires a protein structure,
but does not rely on any prior knowledge of the pores which pass through
the protein.  The method first locates high dielectric spots in the center of 
the protein and then generates swarms of configurationally biased Monte Carlo
trajectories which pass from one side of the membrane to the other.  The trajectories
are driven by both geometric and electrostatic features of the protein.  Once
energetically favorable membrane-spanning trajectories are produced, a subsequent
annealing step based only on geometry locates a unique path through the protein.
Analysis of the radius and potential profiles, along with a listing of pore-lining
residues, yields useful information concerning the functioning of the channel. The method
can be applied to study multiple ion transit pathways for ions of any charge.

The method was successfully tested on the potassium channel. Then, we examined
possible chloride ion pathways on the intracellular side of the bacterial
chloride channel homologues.  The chloride channel is challenging due to multiple
ion transit pathways, the curvature feature of the pores, and possible blockage
of the pores by protein residues. 
The results were compared with recent 
physiological experiments on the ClC-0 channel
which indicated important electrostatic effects near the intracellular entrance
of the selectivity filter.  The TransPath results shed light on the energetics
resulting from a range of mutations on key charged residues.  We are currently
also utilizing the TransPath method in conjunction with molecular dynamics 
simulations to explore pore shape 
and energetic fluctuations induced
by multi-ion passage through the filter of the ClC channel homologues.

\section{Acknowledgements}

We gratefully acknowledge the support of the Department of Defense MURI 
program. We thank John Cuppoletti, Danuta Malinowska, and 
Rob Coalson for many helpful discussions. Z. F. Kuang 
 is grateful to Bob Eisenberg for bringing him to the ion channel 
field. The protein 
structures were displayed by VMD \citep{Humphrey96}.

\newpage

\bibliography{tlbtranspath}

\newpage

\begin{table}
\begin{center}
\vspace{0.5cm}
\label{ta:c}
\small
\begin{tabular}{|c|c|c|c|c|c|c|c|c|}
\hline
\multicolumn{9}{|c|}{$z$ locations of charged sites for 1OTS} \\ \hline 
Residues &  S107     &  R147    & R340     & R120       & K452    & E111       & E113       & E203      \\ \hline
$z_1$       &  -5.49   &  7.86   & 7.85    & -18.94    & -13.50 & -13.79    & -11.08    & -7.93    \\ \hline
$z_2$       &           &  8.80   & 9.17    & -16.94    &         & -15.18    & -10.00    & -9.03     \\ \hline
\multicolumn{9}{|c|}{Distances of key residues from S$_{int}$ and S$_{cen}$ in~{\AA} } \\ \hline 
Residues & &&&& K452    &  E111  & R451   & \\ \hline
S$_{int}$     &&&&  & 6.27   &  6.70, 8.21   & 9.34, 10.29 &    \\ \hline
S$_{cen}$    &&&&  &  11.6    &  12.1, 13.0   &    &  \\ \hline
\end{tabular}
\end{center}
\caption{Above: $z$ locations of charged sites of important 
residues for 1OTS. Below: distances of key residues from the S$_{int}$
and S$_{cen}$ binding sites.}
\end{table}

.
\newpage
\begin{table}
\begin{center}
\vspace{0.5cm}
\label{ta}
\small
\begin{tabular}{|c|c|c|c|c|c|c|c|c|c|}
\hline
\multicolumn{10}{|c|}{Potential and shift at S$_{int}$ for different mutations} \\ \hline 
111/452   &  E-K+ & E0K+ & E-K0 & E0K0 & K+E- & K+E0 & E0E-  & K+K+ &  E-E- \\
 \hline
Potential(V)  & 0.435        & 0.582       & 0.190     & 0.336    & 0.382    & 0.513 & 0.211 &  1.064     &  0.085 \\
\hline
Shift(V)  & 0.000       & 0.147       & -0.245     & -0.099    & -0.053    & 0.078 &  -0.224 &  0.629     &  -0.350 
  \\ \hline

\multicolumn{10}{|c|}{Potential at S$_{int}$ for 1OTS} \\ \hline 

States        &  default   &  default+Cl & E113p     & E113p+Cl &&&&& \\ \hline
Potential(V)  &  0.230   &  0.114    & 0.294   & 0.178  &&&&&  \\ \hline

\multicolumn{10}{|c|}{Potential at S$_{int}$ for 1OTU} \\ \hline 

States        &  default   &  default+Cl & E113p     & E113p+Cl  &&&&&\\ \hline
Potential(V)  &  0.235   &  0.125    & 0.296   & 0.186    &&&&&\\ \hline

\end{tabular}
\end{center}
\caption{Top: potential and shift at S$_{int}$ for different 
mutations of the 111/452 pair in 1OTU. Middle: potential at S$_{int}$ for 1OTS. 
Bottom: potential at S$_{int}$ for 1OTU.} 
\end{table}

.
\newpage
\begin{figure}
\begin{center}
\includegraphics[scale=1.2]{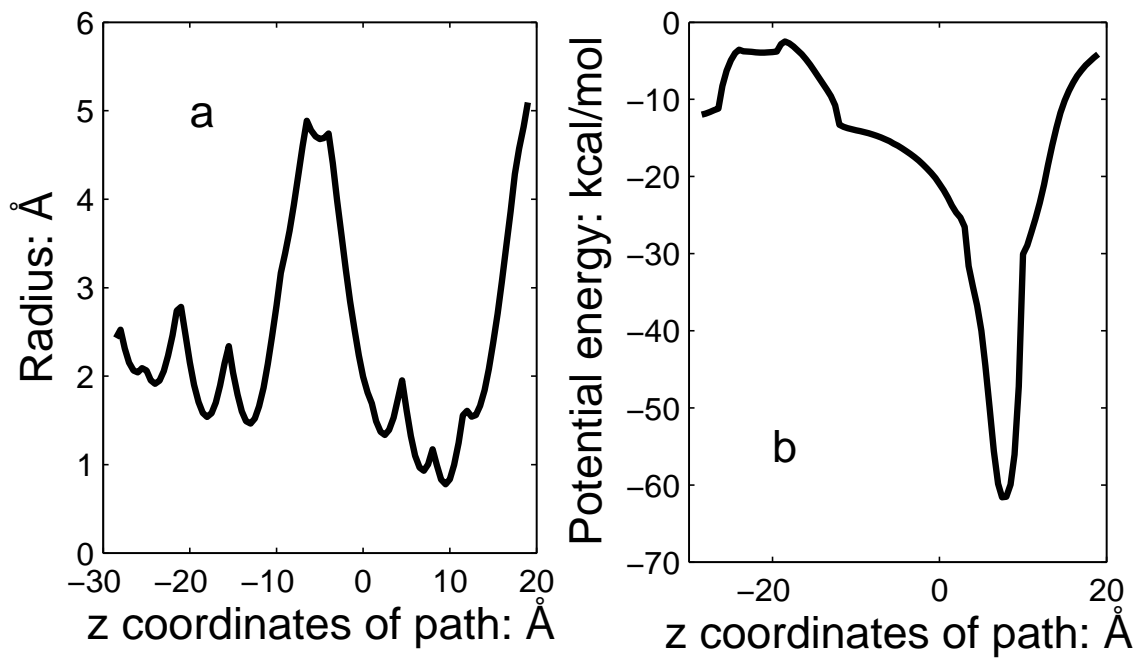}
\end{center}
\caption{(a) The radius profile for the KcsA potassium channel. (b) The 
potential profile.}
\label{fig:1}
\end{figure}

.
\newpage
\begin{figure}
\begin{center}
\includegraphics[scale=1.2]{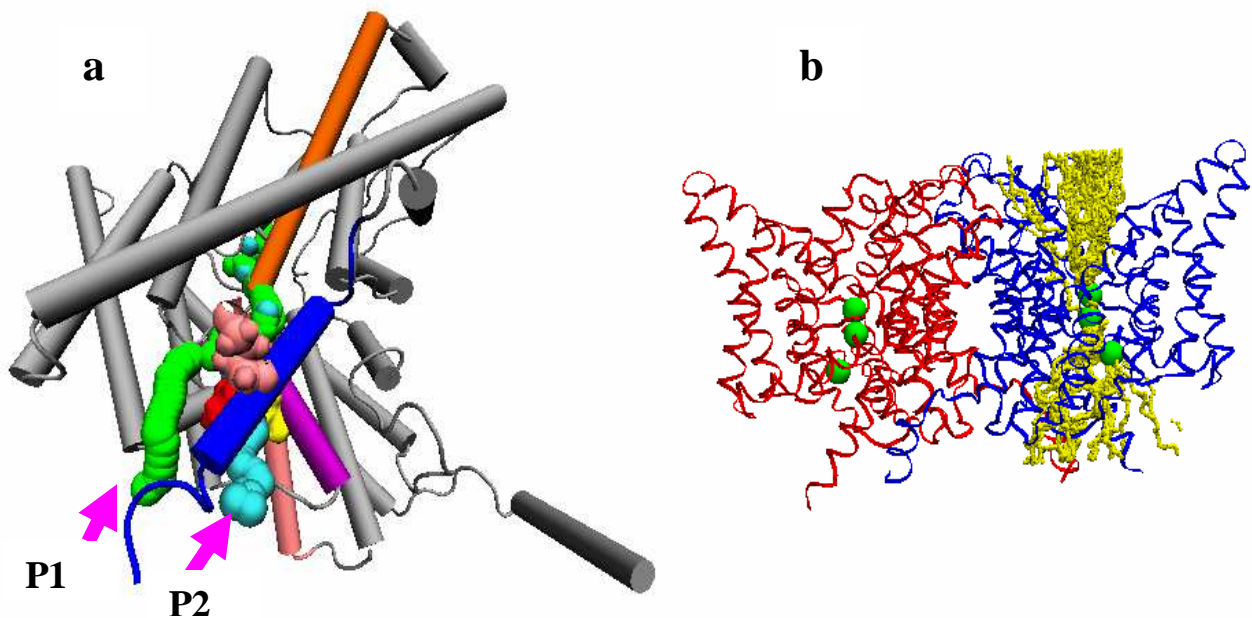}
\end{center}
\caption{(a) The two wild-type 1OTS anion transit pathways P1 and P2, which differ 
in the
intracellular region. The view is from the side (similar to Fig.~2b), 
with the protein slightly rotated 
to display the intracellular vestibule. The P1 path (green) is 
the vestibular path, while the P2
path (cyan) passes near the D and R helices.  Helix R is blue, 
helix D is purple, helix F is peach, and helix N is rust/orange.  Residue E111
is yellow, T452 is red, and R451 is peach, all from the wild-type 1OTS structure.
(b) Display of an ensemble of CBMC
trajectories (yellow) for the mutant 1OTU structure. Notice the spreading in the
two vestibular regions and the constriction through the selectivity
filter. The green spheres are the locations of the three anion binding sites. }
\label{fig:2}
\end{figure}

.
\newpage
\begin{figure}
\begin{center}
\includegraphics[scale=1.2]{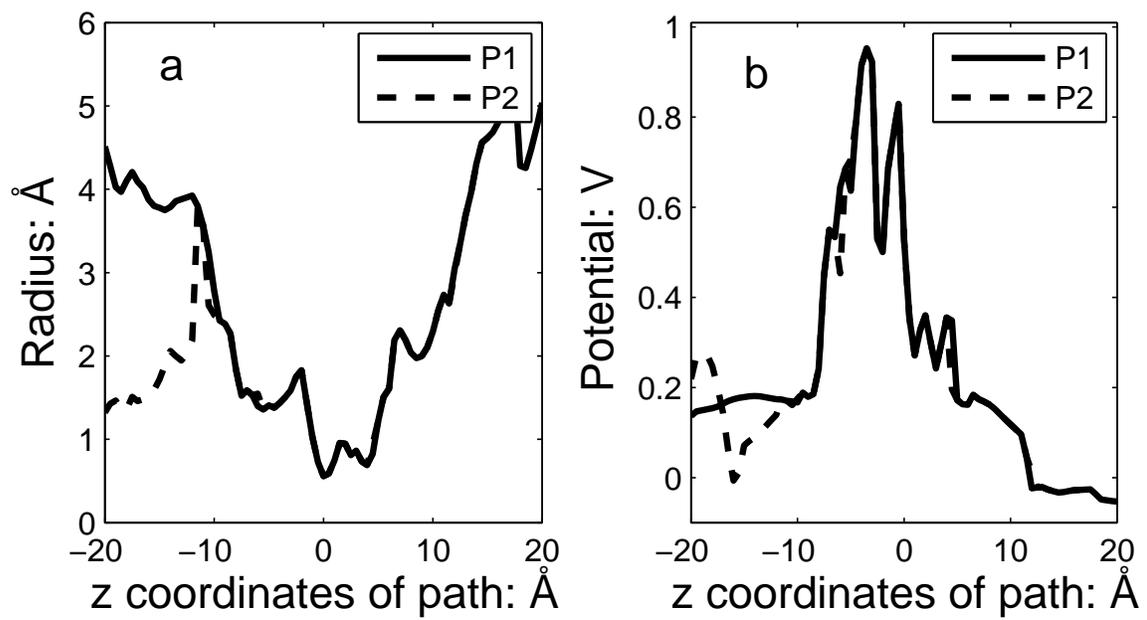}
\end{center}
\caption{(a) The radius profiles for the two paths P1 and P2 in 1OTS. (b) The 
potential profiles.}
\label{fig:3}
\end{figure}


\end{document}